\documentclass[graphicx,superscriptaddress,prl,longbibliography,reprint]{revtex4-1}

\usepackage[version=3]{mhchem} 
\usepackage[T1]{fontenc}
\usepackage{siunitx}
\usepackage{graphicx}

\usepackage[normalem]{ulem}
\useunder{\uline}{\ul}{}
\begin{document}
\author{Robin J. Dolleman}
\affiliation{Kavli Institute of Nanoscience, Delft University of Technology, Lorentzweg 1, 2628CJ, Delft, The Netherlands}
\affiliation{ARC Centre of Excellence in Exciton Science, School of Mathematics and Statistics, The University of Melbourne, Victoria, 3010, Australia}
\affiliation{Present address: Second Institute of Physics, RWTH Aachen University, 52074, Aachen, Germany }
\author{Debadi Chakraborty}
\affiliation{ARC Centre of Excellence in Exciton Science, School of Mathematics and Statistics, The University of Melbourne, Victoria, 3010, Australia} 
\author{Daniel R. Ladiges}
\affiliation{ARC Centre of Excellence in Exciton Science, School of Mathematics and Statistics, The University of Melbourne, Victoria, 3010, Australia}
\affiliation{Center for Computational Sciences and Engineering, Lawrence Berkeley National Laboratory, 1 Cyclotron Rd, Berkeley, CA 94720, USA}
\author{Herre S. J. van der Zant}
\affiliation{Kavli Institute of Nanoscience, Delft University of Technology, Lorentzweg 1, 2628CJ, Delft, The Netherlands} 
\author{John E. Sader}
\affiliation{ARC Centre of Excellence in Exciton Science, School of Mathematics and Statistics, The University of Melbourne, Victoria, 3010, Australia} 
\author{Peter G. Steeneken}
\email{P.G.Steeneken@tudelft.nl}
\affiliation{Kavli Institute of Nanoscience, Delft University of Technology, Lorentzweg 1, 2628CJ, Delft, The Netherlands} 
\affiliation{Department of Precision and Microsystems Engineering, Delft University of Technology, Mekelweg 2, 2628 CD, Delft, The Netherlands}

\title{Squeeze--film effect on atomically thin resonators in the high--pressure limit}

\begin{abstract}
The resonance frequency of membranes depends on the gas pressure due to the squeeze-film effect, induced by the compression of a thin gas film that is trapped underneath the resonator by the high frequency motion. This effect is particularly large in low-mass graphene membranes, which makes them promising candidates for pressure sensing applications. Here, we study the squeeze-film effect in single layer graphene resonators and find that their resonance frequency is lower than expected from models assuming ideal compression. To understand this deviation, we perform Boltzmann and continuum finite-element simulations, and propose an improved model that includes the effects of gas leakage and can account for the observed pressure dependence of the resonance frequency. Thus, this work provides further understanding of the squeeze-film effect and provides further directions into optimizing the design of squeeze-film pressure sensors from 2D materials. 
\end{abstract}
\maketitle

The hexagonal structure of graphene gives rise to unique electronic properties, which has attracted considerable interest in the scientific community \cite{geim2007rise}. Moreover, the strong bonds between the carbon atoms make this material one of the strongest materials ever measured \cite{lee2008measurement,wang2017single}. These properties, in combination with a low mass per area, high-flexibility, and gas impermeability, make graphene an interesting material for sensing applications \cite{bunch2008impermeable,koenig2012selective,smith2013electromechanical,dolleman2016graphene,davidovikj2017static,vollebregt2017suspended,dolleman2015graphene,verbiest2018detecting,fan2019graphene,fan2019suspended,wittmann2019graphene,blaikie2019fast,dolleman2020semi,roslon2020high,vsivskins2020sensitive,lemme2020nanoelectromechanical}. Graphene pressure sensors using the squeeze-film effect promise high responsivity, while at the same time considerably reducing the sensor area, compared to state-of-the art sensors \cite{vollebregt2017suspended,dolleman2015graphene}. Squeeze-film pressure sensors compress a gas in a shallow cavity underneath the vibrating membrane, which raises the stiffness of the system by an amount that depends on the gas pressure \cite{steeneken2004dynamics,bao2007squeeze,andrews1993comparison,andrews1993resonant,southworth2009pressure,kumar2015mems,naserbakht2019squeeze,dantan2020membrane}. If the membrane vibrates at sufficiently high frequencies and if the compression is isothermal, the resonance frequency, $\omega$, of the graphene membrane has a pressure dependence that can be described by \cite{dolleman2015graphene}:
\begin{equation}\label{eq:sqz}
\omega^2 = \omega_0^2 + \frac{p}{g_0 \rho h},
\end{equation}
where $\omega_0$ is the mechanical resonance frequency in vacuum, $p$ the ambient gas pressure, $g_0$ the distance between the moving membrane and the fixed substrate and $\rho h$ is the membrane's mass per square meter. Note, that due to the ultralow $\rho h$ of single--layer graphene, large shifts in frequency $\omega$ can be expected from Eq. \ref{eq:sqz}. Although Eq. \ref{eq:sqz} describes the pressure dependence of the resonance frequency well at low pressures, significant deviations from this equation are observed at pressures above 100 mbar, which are presently not understood \cite{dolleman2015graphene}. To clarify these observations, a more in-depth understanding of the squeeze-film effect in these nanoscale systems is required.

  Here, we study the squeeze-film effect on single-layer graphene membrane resonators as a function of pressure for different types of gases. A clear, gas dependent, deviation of the frequency response from the behaviour predicted by Eq. \ref{eq:sqz} is found in the 100--1000 mbar pressure range. Both the experiments and simulations show that the onset of the deviation is correlated to the quality factor of the resonance, which can be accounted for by the gas leakage out of the gap region with a rate that is characterized by a single relaxation time. This work therefore provides a deeper understanding on squeeze-film dynamics and its effect on atomically thin membranes. The results can contribute to improving the operation and design of squeeze-film pressure sensors.

\begin{figure*}
\includegraphics{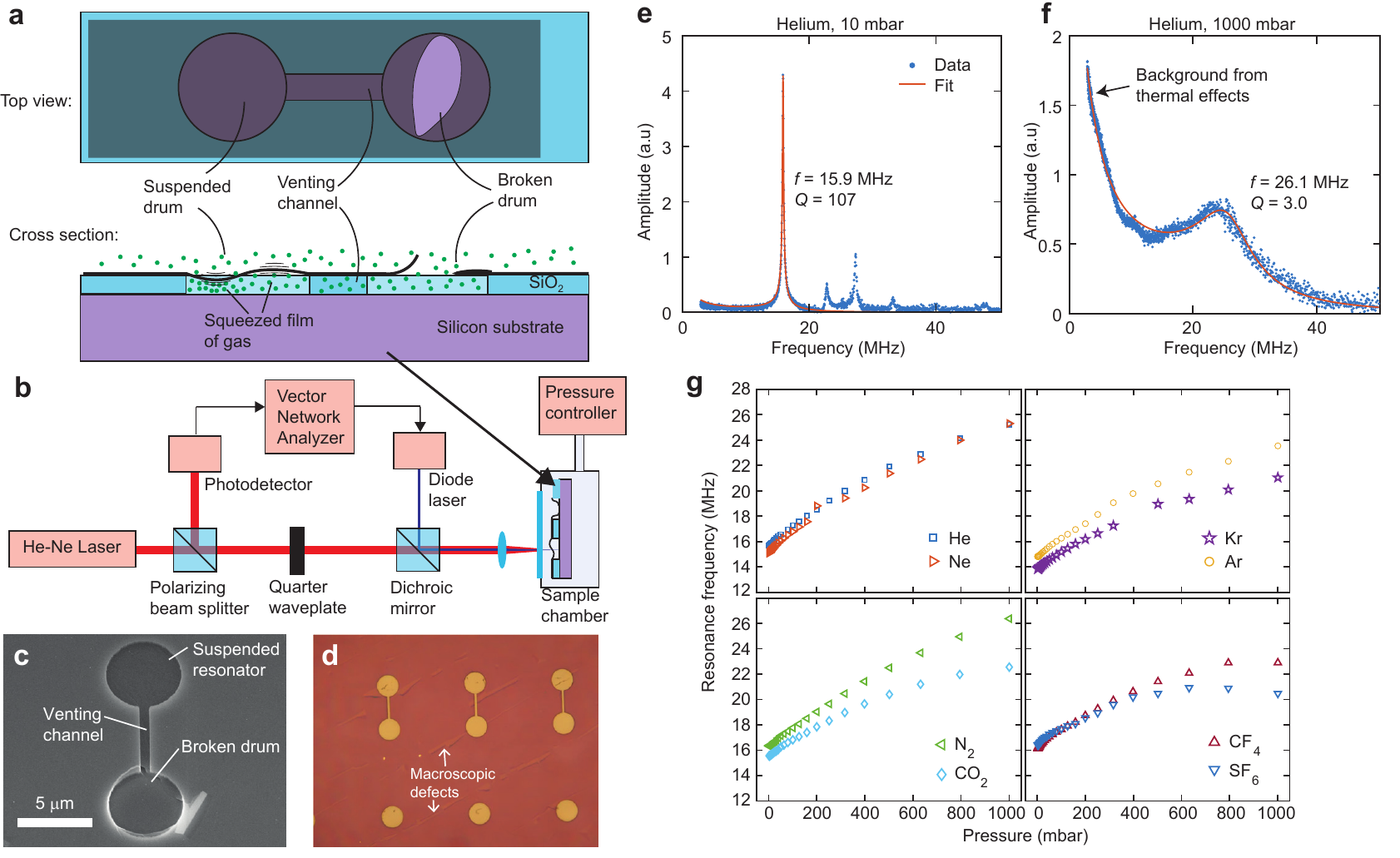}
\caption{Graphene samples used in the experiment and the experimental setup to actuate and detect their motion. \textbf{a} Top-view and cross-section of a graphene resonator on a dumbbell-shaped cavity. The cross-section shows a schematic drawing of the resonator's motion, highlighting where compression of the gas occurs due to the squeeze-film effect.  \textbf{b} Interferometric setup used to actuate and detect the motion. \textbf{c} Scanning electron microscope image of a representative device used in this study; the bottom half of the dumbbell is broken while the top half is whole. \textbf{d} Optical microscope image of a part of the chip, showing the coverage of graphene over the chip's surface and the macroscopic defects which helps one part of the dumbbell to break.  \textbf{e} Measured amplitude of motion for a 5-micron diameter membrane as a function of frequency in a 10-mbar helium environment, and a fit (red lines) to the first resonance peak and the low-frequency background using the procedure described in the Supporting information S1. The measured amplitude is corrected for any delays in the electronic components of the experimental setup.  \textbf{f} Same as \textbf{e}, but at a higher pressure of 1000 mbar. The background at low frequencies originates due to thermal effects, as explained in the Supporting information S1. \textbf{g} Pressure-dependent resonance frequency for 8 different gases measured on a 5--micron diameter drum. \label{fig:setup}}
\end{figure*}

 The samples consist of suspended single-layer graphene grown by chemical vapour deposition which are suspended over a dumbbell shaped cavity (Fig. \ref{fig:setup}\textbf{a}, \textbf{c--d}) and of which the graphene over one half of the dumbbell is broken. Since the other half of the graphene dumbbell remains intact, a venting channel is created towards the environment and this prevents pressure differences from forming across the membrane's surface, thus preventing resonance frequency shifts due to pressure induced membrane tension \cite{bunch2008impermeable,dolleman2015graphene,dolleman2016graphene}. The venting channel therefore ensures that any frequency shift can be attributed to the squeeze film effect. The fabrication process to produce these samples is identical to that reported in several previous works \cite{dolleman2017optomechanics,dolleman2017amplitude,dolleman2018opto,dolleman2019high,dolleman2020nonequilibrium} and described in more detail in the Supporting information S1. The sample is mounted in a chamber which can contain different gases at a controlled pressure. The membrane is opto-thermally actuated, and its motion is read out using the optical setup shown in Fig. \ref{fig:setup}\textbf{b}. By fitting to the frequency responses, the resonance frequency and the quality factor is extracted as a function of pressure (Fig. \ref{fig:setup}\textbf{e-f}) More details regarding the samples, experimental setup and data analysis can be found in the Supporting information S1.  

Figure \ref{fig:setup}\textbf{g} shows the pressure-dependent resonance frequency for 8 different gases on a 5-micron diameter drum. For all gases, the resonance frequency increases as a function of pressure due to the squeeze-film effect. It is observed that the total frequency shift depends on the type of gas: gases with a high molecular mass such as CF$_4$ and SF$_6$ show a lower shift than gases with a low molecular mass such as He and Ne. Furthermore, whilst most gases show a monotonic increase of the resonance frequency with pressure, SF$_6$ also shows a decrease in resonance frequency at high pressures. Both the gas dependence and the decrease in resonance frequency are not in agreement with Eq. \ref{eq:sqz} and we will study this in more detail. 
 
\begin{figure}
\includegraphics{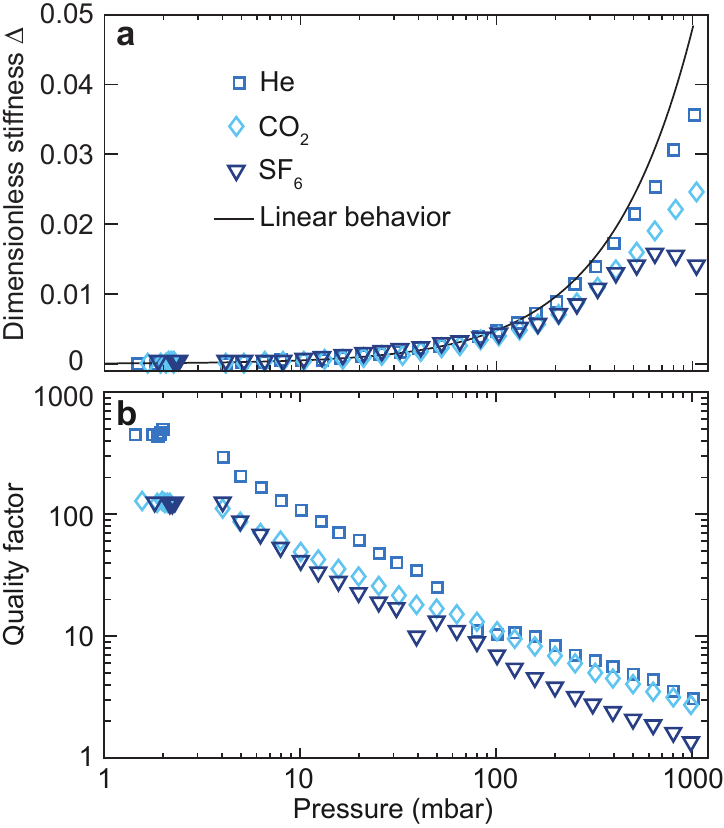}
\caption{Measured squeeze-film stiffness and quality factor as function of pressure and gas for a 5-micron diameter single-layer graphene drum. \textbf{a} Dimensionless stiffness $\Delta$ as function of gas pressure for 3 different gases. The linear behaviour line in both figures is obtained by fitting a polynomial to the helium response and plotting only the linear part. \textbf{b} Quality factor of the fundamental resonance as function of pressure for 3 different gases.  \label{fig:results-shift}}
\end{figure}
Since the vacuum resonance frequency ($\omega_0$) can change somewhat between consecutive measurements (Fig. \ref{fig:setup}\textbf{g}), it is convenient to rescale the measured resonance frequency as a function of pressure. We therefore define a dimensionless squeeze-film stiffness $\Delta$, that rescales the measured pressure-dependent resonance frequency $\omega$, based on Eq. \ref{eq:sqz}:
\begin{equation}
\Delta = (\omega^2 - \omega_0^2) \frac{\rho h g_0 }{p_{\mathrm{ref}}},
\end{equation}
where $p_{\mathrm{ref}}$ is a reference pressure chosen to be 1000 mbar, $\rho h$ for single-layer graphene is $\num{7.7e-7}$ kg/m$^2$ and $g_0$ is measured to be 300 nm. This dimensionless stiffness $\Delta$ allows us to compare the squeeze-film effect for different measurements, even when the vacuum resonance frequency of the resonator has shifted. 

Figure \ref{fig:results-shift}\textbf{a} shows the measured dimensionless squeeze-film stiffness $\Delta$ as a function of pressure for three different gases from the same drum as Fig. \ref{fig:setup}\textbf{g}, the results on 5 other gases are shown in the Supporting information S2.
To analyse the deviations from the linear behaviour in Fig. \ref{fig:results-shift}\textbf{a}, we fit a polynomial to the $\Delta$ of helium, $\Delta =  A_1 p + A_2 p^2 + ...$, where the order is increased until we obtain a reasonable fit in the whole pressure range. In this case, the fit was made using a second order polynomial. For the squeeze-film effect in the ideal case, meaning described well by Eq. \ref{eq:sqz}, we expect that $\Delta = \frac{p}{p_{\rm ref}}$. However, since the mass of the single layer graphene can deviate significantly from theory \cite{dolleman2019mass,steeneken2021dynamics} and is unknown to us, we assume that in the ideal case $\Delta$ scales linearly with pressure $p$.  This behaviour is represented by the black line in Fig. \ref{fig:results-shift}\textbf{a}, where we only plot $\Delta_{\rm lin} =  A_1 p$. For all the gases  in Fig. \ref{fig:results-shift}\textbf{a}, we observe that $\Delta$ is similar up to 100 mbar and described well by this linear behaviour. However, at higher pressures, a significant gas dependence of $\Delta$ is observed and all gases show a lower stiffness than expected. 

The quality factor as a function of gas pressure is shown in Fig. \ref{fig:results-shift}\textbf{b}. For all gases, the quality factor decreases as a function of pressure due to the viscous dissipation forces, with a similar slope on a logarithmic scale. Throughout the pressure range studied here, low-density gases tend to show less dissipation than high-density gases. At any pressure, the quality factor changes with less than an order of magnitude as a function of the molecular weight of the gases.

\begin{figure*}
\includegraphics{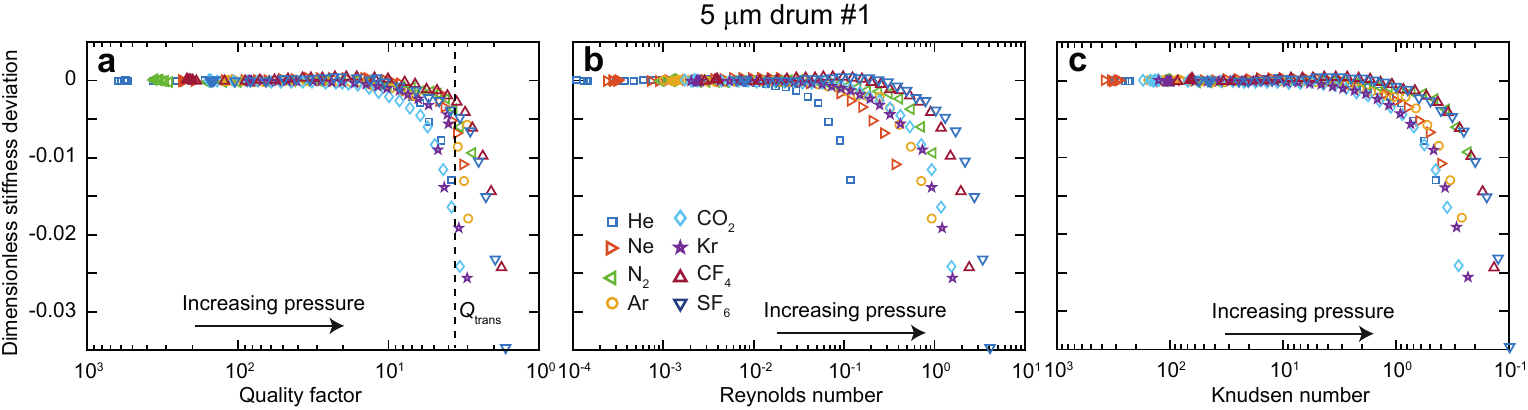}
\caption{Dimensionless analysis on the experimental data, showing the stiffness deviation with respect to the linear part of helium's dimensionless stiffness. \textbf{a} Dimensionless stiffness deviation measured on a 5-micron diameter drum as a function of quality factor for 8 different gases, \textbf{b} as a function of Reynolds number and \textbf{c} as a function of Knudsen number.  \label{fig:dimensionless}}
\end{figure*}
To examine the cause of the reduced stiffness at high pressures, we perform a dimensionless number analysis. For each gas, we define the dimensionless stiffness deviation as $\Delta_{\rm dev} = \Delta-\Delta_{\mathrm{lin}}$. The dimensionless stiffness deviation is then compared to other dimensionless numbers. First, is the quality factor of resonance. Second is the Reynolds number, that gives the ratio between inertial and viscous forces. For an oscillating flow this is defined as \cite{bao2007squeeze}:
\begin{equation}
\mathrm{Re} = \frac{\omega \rho_g g_0^2}{\mu}
\end{equation}
where $\rho_g$ is the density of the gas and $\mu$ the viscosity. The third dimensionless number is the Knudsen number, which characterizes the degree of gas rarefaction and is defined as:
\begin{equation}
\mathrm{Kn} = \frac{\lambda}{g_0},
\end{equation}
 where $\lambda$ is the mean free path of the molecules of the gas. 
 
 Figure \ref{fig:dimensionless} shows the result of the dimensionless analysis for a 5-micron diameter drum and a 4-micron diameter drum for all eight gases used in this study. Additional datasets with different drums are shown in the Supporting information S2.  For easier comparison, the horizontal axis of the quality factor and Knudsen number are reciprocal such that the left--hand side of the graph corresponds to low pressure and the right--hand side to high pressure. In all graphs, the stiffness deviation is initially zero and then goes towards negative values, this corresponds to the deviations from linear behaviour discussed above. 

To find the potential cause of the deviation from linear behaviour, we look at the transition point where the stiffness deviation is nonzero. 
We take the values of $Q$, $\mathrm{Re}$ and $\mathrm{Kn}$ at $\Delta_{\rm dev,trans} = -0.005$ to define this transition point. Determining the transition point for each gas and each drum used in this study (including those presented in the Supporting information S2), we find that the transition occurs at $Q = 3.81 \pm 1.07$, $\mathrm{Re} = 0.55 \pm 0.38$ and $\mathrm{Kn} = 0.82 \pm 0.55$, where the error bars represent the standard deviation. The relative standard deviation $\sigma$ for each parameter is: $\sigma_Q = 28\%$, $\sigma_{\mathrm{Re}} = 50 \%$ and $\sigma_{\mathrm{Kn}} = 69\%$. This analysis shows that transition point where the stiffness deviates from the ideal linear behaviour is thus most strongly associated with a certain number of the quality factor ($Q_{\rm trans} \approx 3.8$), since the this shows the smallest relative spread compared to the Reynolds and Knudsen number. 

The experimental analysis reveals that Eq. \ref{eq:sqz} is no longer valid when the quality factor is lower than approximately 3.8. However, due to experimental uncertainties, we cannot fully rule out that this effect still originates from transitions in the fluid flow characterized by the Reynolds and Knudsen number. Therefore, we use simulations using two approaches and analyse them in a similar manner as the experimental data. The first approach employs finite element simulations by solving the compressible unsteady Stokes equation for the gas together with Navier's equation for the solid membrane using the eigenfrequency solver of COMSOL Multiphysics \cite{dc2013jpcc, dc2015pof}. An axisymmetric slice of the drum is considered  in this continuum model (see Supporting information S3). No-slip boundaries were implemented on the solid surfaces. The second approach includes the effects of gas rarefaction by simulating the system using the frequency-domain Monte Carlo Method \cite{ladiges2015fdmcA, ladiges2015fdmcB}. This approach solves the Boltzmann Transport Equation (BTE), and should, therefore, provide more accurate results for low gas pressures. The simulations are benchmarked against a previously published dataset \cite{dolleman2015graphene} from a 5-micron diameter drum from 31-layer graphene over a 400 nm deep cavity, which is shown in the Supporting information S3.  While both the BTE and continuum simulations show the experimentally observed deviations in the stiffness, the BTE simulations are computationally expensive at high pressures. We therefore choose to perform the dimensionless analysis with different gases on the continuum simulations, which are easier to perform at high pressures with different gases.
 
  \begin{figure*}
\centering
\includegraphics{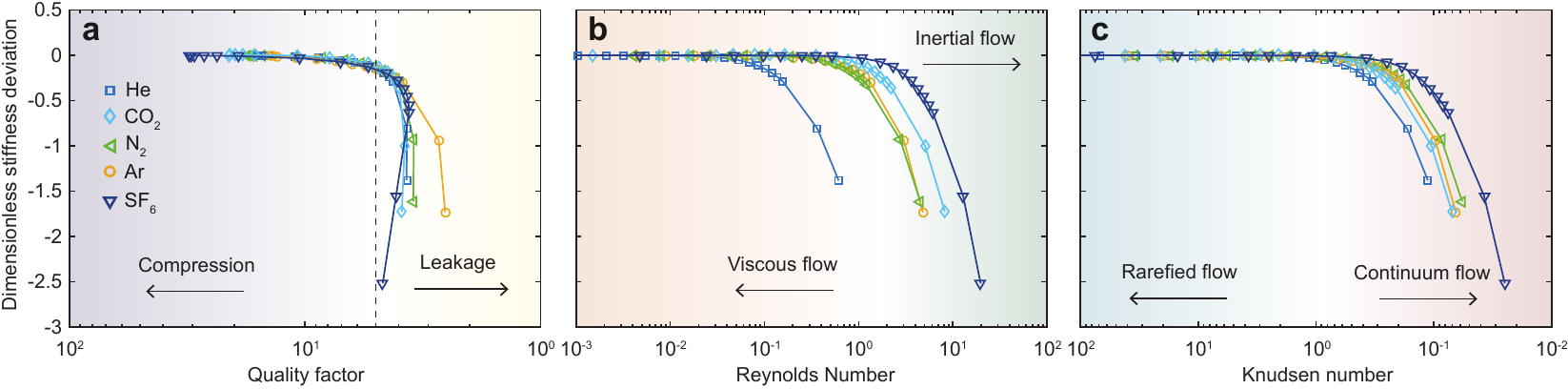}
\caption{Dimensionless analysis of the simulated graphene drum. Dimensionless stiffness deviation as a function of  (\textbf{a}) the mechanical quality factor, (\textbf{b}) the Reynolds number and (\textbf{c}) the Knudsen number. \label{fig:dimensionlesssim}}
\end{figure*}
 The dimensionless analysis on the simulated data is shown in Fig. \ref{fig:dimensionlesssim}. This reveals that there is a single quality factor of $Q \approx 5$ where the transition takes place indicated by the dashed line in Fig. \ref{fig:dimensionlesssim}. The simulations also show that the point $\Delta_{\rm dev} = -0.005$ is not at the same Reynolds or Knudsen number, confirming that the Q-factor is the strongest correlated to the point beyond  which deviations from Eq. \ref{eq:sqz} start to increase. Comparing the dimensionless stiffness deviation shift to the case of the 5-micron diameter drum in the experiments (Fig. \ref{fig:dimensionless}), the deviation found in the simulation are similar as function of Reynolds number and Knudsen number, with transitions taking place in similar ranges. Note, that the continuum simulations are only accurate in the near-continuum regime where $\mathrm{Kn} < 1$. Since the deviations from linear behaviour are indeed found in this regime, we can use these simulations to study them, but the model is not accurate at low pressures where $\mathrm{Kn} > 1$.

The experiments and simulations thus both show a smaller squeeze-film stiffness effect compared to Eq. \ref{eq:sqz}, and the transition is point is association with a certain number of the quality factor. To investigate why this happens, we construct a simple one-dimensional model for the motion of the membrane in the frequency domain:
 \begin{equation}\label{eq:motion}
 -\omega^2  x +  \omega_0^2 x =  \beta \Delta p,
 \end{equation}
where $\beta$ is a proportionality constant and the change in gas pressure $\Delta p$ is related to the displacement $x$ as \cite{roslon2020high}:
  \begin{equation}\label{eq:pressurefield}
  i \omega \Delta p + \frac{1}{\tau_g} \Delta p = \gamma i \omega x,
 \end{equation}
 where $\gamma$ is another proportionality constant. This equation is equivalent to the linearized Reynolds equation \cite{bao2007squeeze}, where the lateral position dependence of the pressure has been projected onto one generalized coordinate $\Delta p$. The relaxation of $\Delta p$ after the membrane compresses the gas is approximated with a single characteristic leak time constant $\tau_g$. Solving this coupled system (Eqs. \ref{eq:motion}--\ref{eq:pressurefield}) gives complex eigenvalues (see Supporting information S4):
 \begin{equation}\label{eq:sqznew}
\omega^2 = \omega_0^2  + \frac{p}{g_0 \rho h}  \frac{\sigma^2}{\sigma^2 +1}+   i\frac{p}{g_0 \rho h}  \frac{\sigma}{\sigma^2 +1} ,
\end{equation}
where $\sigma = \tau_g \omega$ is the dimensionless squeeze number, which compares the timescale of compression $1/\omega$ to the pressure relaxation time $\tau_g$. For a squeeze number $\sigma \gg 1$, Eq. \ref{eq:sqznew} becomes equal to Eq. \ref{eq:sqz}. However, for lower values of sigma, the resonance frequency (as determined by the real part of Eq. \ref{eq:sqznew}) becomes smaller than that in Eq. \ref{eq:sqz}. 
Using that the quality factor $Q_f = \mathcal{R}(\omega)/2\mathcal{I}(\omega)$, we find (see Supporting Information S4):
\begin{equation}\label{eq:Qfmain}
Q_f = \frac{\sqrt{\xi^2 + 1} + 1}{2 \xi},
\end{equation}
where
\begin{equation}\label{eq:xQf}
\xi = \frac{\sigma }{\frac{\omega_0^2}{\omega_{\mathrm{sqz}}^2} (\sigma^2 + 1) + \sigma^2 },
\end{equation}
where $\omega_{\mathrm{sqz}}^2 = \frac{p_a}{g_0 \rho h}$. The quality factor thus depends on two dimensionless numbers: the dimensionless squeeze number $\sigma$ and the relative shift in resonance frequency $\omega_{\mathrm{sqz}}/\omega_0$. 

  \begin{figure}
 \includegraphics{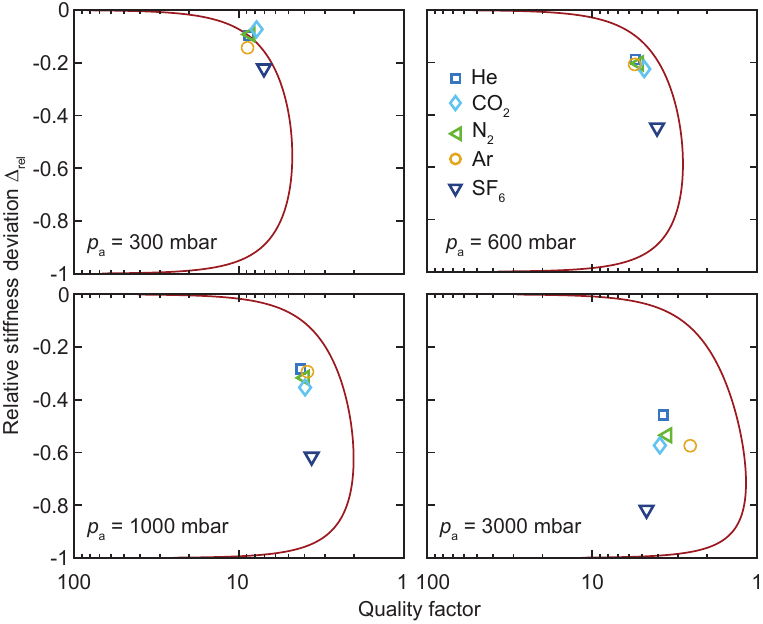}
 \caption{Relative stiffness deviation (Eq. \ref{eq:deltarel}) versus the quality factor from the continuum model. The red lines are predictions at fixed pressures obtained from the single relaxation time models from Eq. \ref{eq:sqznew}.   \label{fig:predictomega}}
 \end{figure}
Equation \ref{eq:sqznew} predicts a relation between the stiffness and damping due to the squeeze-film effect for a certain fixed $\sigma$. To verify this relation, we define the relative stiffness deviation, which can be related to the dimensionless squeeze number $\sigma$ as shown in the Supporting information S5:
\begin{equation}\label{eq:deltarel}
\Delta_{\rm rel} = \frac{\Delta - \Delta_{\rm lin}}{\Delta_{\rm lin}}
\end{equation}
This equation is inverted to obtain $\sigma$ and then substituted into Eqs. \ref{eq:Qfmain}--\ref{eq:xQf}, which results in a relationship between the quality factor $Q_f$ and relative stiffness deviation $\Delta_{\rm rel}$ where the only free parameter is the pressure $p_a$ (assuming $\rho h$ and $g_0$ are known). Figure \ref{fig:predictomega} shows this relationship at fixed pressures compared to the results of the continuum simulation. The trend predicted by Eqs. \ref{eq:Qfmain}--\ref{eq:deltarel} is in qualitative agreement with the results from the continuum simulation at pressures of 300 and 600 mbar. 
However, at 1000 and 3000 mbar, the single relaxation time model predicts lower quality factors than those obtained by the continuum model. The simplifications behind the single relaxation time model may account for this difference, since the continuum model can take multiple timescales into account.

From the model we conclude that the cause of the deviations from Eq. \ref{eq:sqz} is the effect of gas leakage, leading to relaxation times that are not long enough compared to the period of oscillation to be neglected. These effects modify the dimensionless squeeze number and reduce the resonance frequency with respect to Eq. \ref{eq:sqz}.
This reduced responsivity can be mitigated by either increasing the resonance frequency or the leak time constant $\tau_g$. One approach to achieve a higher resonance frequency may be to clean the graphene membrane to reduce the mass. In the Supporting information S3 we show both continuum and BTE simulation results for a helium atmosphere in the case of clean single layer graphene (with a mass of $\num{7.7e-7}$ kg/m$^2$) and a 31--layer device in nitrogen presented in Ref. \cite{dolleman2015graphene}. These simulations suggest that if the graphene is cleaned in order to reduce the mass, the responsivity of the squeeze-film pressure sensors will significantly improve. We attempted to clean the single-layer graphene in this study by argon annealing at 400$^{\circ}$C, however this resulted in destruction of the drums. 

In the future, Eq. \ref{eq:sqznew} can be used instead of Eq. \ref{eq:sqz} to understand the effects of gas leakage on the responsivity of squeeze-film pressure sensors.  Their behaviour crucially depends on the exact value of $\tau_g$, which is determined by the exact geometry of the device and the properties of the gas flow. Therefore, the simulations on a device level, such as the continuum and BTE simulations presented in this work, are crucial to understand the behaviour of these sensors. 

In conclusion, we study the squeeze-film effect in single-layer graphene resonators in different gases and find a lower resonance frequency than expected. By examining the experiments and simulations of the device we find that the transition towards this lower resonance frequency regime is correlated to the quality factor of the resonance. We explain this by a one-dimensional mechanical model and show that this is related to a low value of the dimensionless squeeze number. This suggest that, to improve future squeeze-film pressure sensors from graphene, the resonance frequency and the leak time constant need to be increased. The experiments and simulations thus provide an improved understanding into the fluid-structure interaction at the nanoscale and provides a means to better engineer future sensors from suspended 2D materials.

\section*{Supporting information}
The reader is referred to the Supporting information S1 for details on the fabrication, experimental setup and data analysis; this section also includes Refs. \onlinecite{castellanos2013single,dolleman2018transient}. Supporting information S2 shows the remainder of the data in Fig. 2, the resonance frequency and Q factor of 6 more samples, and the dimensionless number analysis on these samples. Supporting information S3 contains further details on the numerical simulations and also includes Refs. \onlinecite{Bird1994,bhatnagar1954model}. Supporting information S4 shows the derivation of the relation between the Q-factor and the dimensionless squeeze number (Eqs. \ref{eq:Qfmain}--\ref{eq:xQf}) and Supporting information S5 further derives the relation between the relative stiffness deviation (Eq. \ref{eq:deltarel}) and the Q-factor.

\section*{Acknowledgments}
\begin{acknowledgements}
The authors thank Applied Nanolayers B.V. for the supply and transfer of the single-layer graphene used in this study.
This work is part of the research programme Integrated Graphene Pressure Sensors (IGPS) with project number 13307 which is financed by the Netherlands Organisation for Scientific Research (NWO).
The research leading to these results also received funding from the European Union's Horizon 2020 research and innovation programme under grant agreement No 785219 and 881603 Graphene Flagship. 
This work has received funding from the EMPIR programme co-financed by the Participating States and from the European Union’s Horizon 2020 research and innovation programme.
 R.J.D. also acknowledges funding from the Mobility Grant within the European Union's Horizon 2020 research and innovation programme under grant agreement No 785219 Graphene Flagship; and support from the University of Melbourne and the Graduate Union of the University of Melbourne inc. 
D.C., D.R.L. and J.E.S. gratefully acknowledge support from the Australian Research Council Centre of Excellence in Exciton Science (CE170100026) and the Australian Research Council Grants Scheme.
D.R.L. also acknowledges funding from the US Department of Energy, Office of Science, Office of Advanced Scientific Computing Research, Applied Mathematics Program under Contract No. DE-AC02-05CH11231.
\end{acknowledgements}

\newpage
~
\appendix
\newpage
\onecolumngrid
\section*{Supporting information}
\subsection*{S1: Methods.}
Fabrication of the samples starts on silicon dies with $\sim$285 nm of silicon dioxide. Dumbbell-shaped cavities are etched in the silicon dioxide layer, with a total depth of approximately 300 nm. Single layer graphene grown by chemical vapor deposition was transferred on top of this substrate using a supporting polymer. This polymer is dissolved, and the sample is subsequently dried using CO$_2$ critical point drying. Besides removing the polymer, this process is used to break the weakest part of the dumbbell, leaving a suspended circular membrane with a venting channel to the environment on the other side (Figs. 1 \textbf{a}, \textbf{c}-\textbf{e} in the main text). The venting channel prevents that pressure differences across the membrane can alter the tension and resonance frequency of the membrane, and thus it ensures that any observed frequency shifts can be attributed to the squeeze-film effect.

Figure 1\textbf{b} in the main text shows the experimental setup to actuate and detect graphene's motion in a controlled gaseous environment. The sample is mounted in a vacuum chamber, which is carefully tested for leaks to ensure the gas composition inside the chamber. A voltage-controlled dual-valve pressure controller is used to control the pressure in this chamber. Eight different gases can be connected to the input of the controller to select the type of gas. A red helium-neon laser is used to read-out the membrane motion by Fabry-Perot interferometry \cite{castellanos2013single,dolleman2017amplitude}. The silicon substrate acts as a fixed mirror, while the suspended graphene membrane acts as the moving mirror. A blue diode laser is used to actuate the motion of the membrane by opto-thermally heating the membrane, which will experience a force due to thermal expansion \cite{dolleman2017optomechanics,dolleman2018opto}.

The squeeze-film effect is measured by characterizing the membrane's amplitude of motion as a function of the frequency of the opto-thermal actuation, which is repeated at different pressures set by the pressure controller. To correct for any frequency dependence arising from the electronic components in the setup, the response of the setup is measured when the blue laser is directly illuminating the photodetector. This measurement is then used to deconvolve the measured response, two examples of such corrected responses are shown in Figs. 1\textbf{e} and \textbf{f} in the main text. To fit a harmonic oscillator response to the data, we must consider the thermal delay, which causes the actuation force to become frequency dependent \cite{dolleman2017optomechanics,dolleman2018transient,dolleman2020nonequilibrium}. A further frequency dependence can emerge from gas leakage, which has an identical frequency dependence \cite{roslon2020high}. This frequency dependence is corrected by fitting the response with a single thermal time constant model:
\begin{equation}\label{eq:fittau}
x = \frac{A}{\omega^2 \tau^2 + 1} - \frac{i A \omega \tau}{\omega^2 \tau^2 +1}, 
\end{equation}
where $A$ is used as a fitting variable. We fit to the imaginary part of the data to extract $A$ and $\tau$; and use those parameters to again deconvolve the data with Eq. \ref{eq:fittau}. In these measurements, usually either thermal or gas leakage effects dominate the response, and a single relaxation time model provided a good fit. However, at some pressures both thermal and gas delay effects occur at the same time, and a single relaxation time does not provide a good fit. In those cases, no frequency and quality factor are fit, and those data points are omitted. After the correction for the frequency dependence of the actuation force, the data is fit using a simple harmonic oscillator model without an additional background. In the example traces in Figs. 1\textbf{e} and \textbf{f} in the main text, the fits to the harmonic oscillator are again multiplied with the fit to frequency-dependence of the actuation force at low frequencies. This shows that the fitting procedure can accurately represent the resonance peak and the background. If the frequency dependence of the actuation force is not considered, the background will cause the fitting to underestimate the resonance frequency when the quality factor of the resonance is low. 

\subsection*{S2: Additional measurements} 
\begin{figure*}[h!]
\includegraphics[scale = 1]{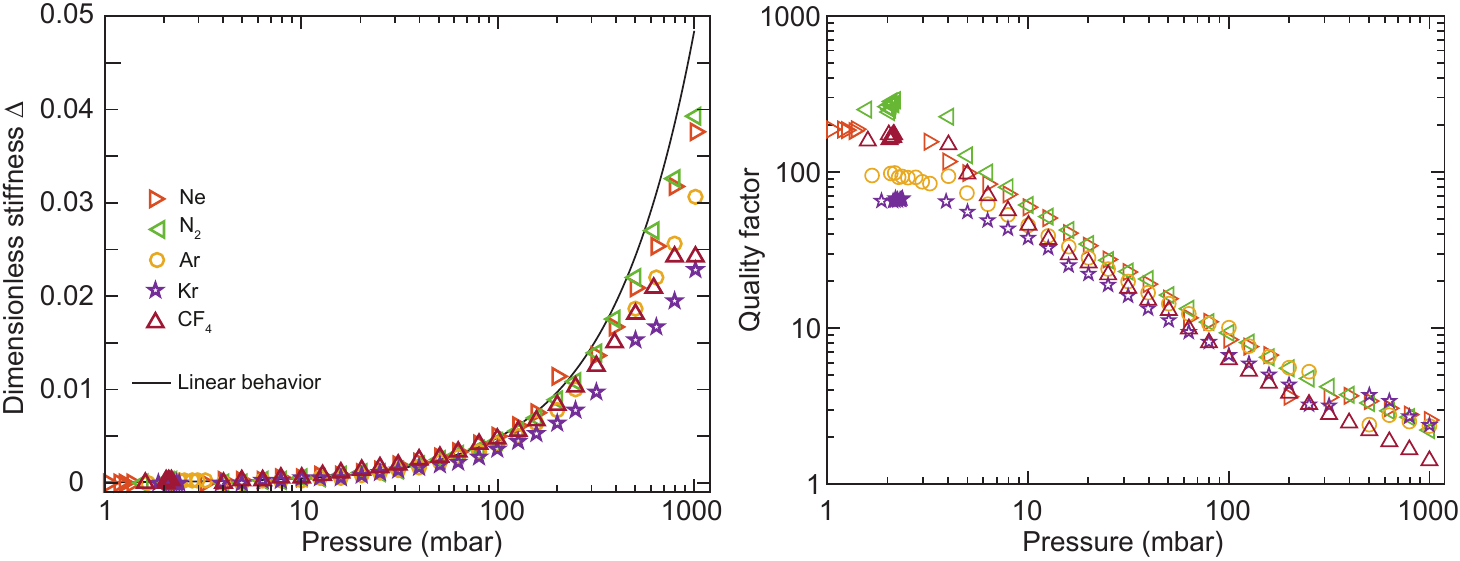}
\caption{Remainder of the dataset in Fig. 2 in the main text.  }
\end{figure*}
\begin{figure*}[h!]
\includegraphics[scale = 0.75]{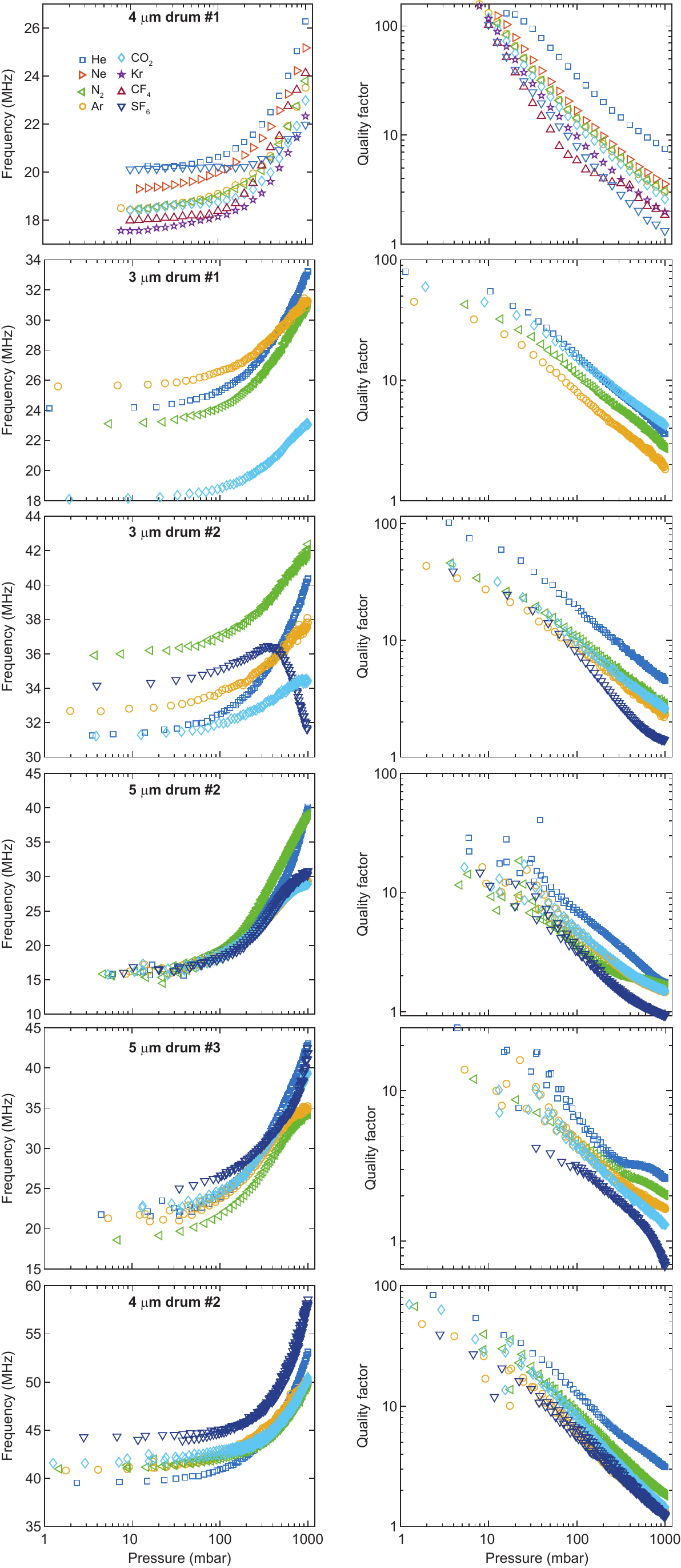}
\caption{Resonance frequency and Q-factor of 6 more samples not shown in the main text.  }
\end{figure*}
\begin{figure*}[h!]
\includegraphics[scale = 0.8]{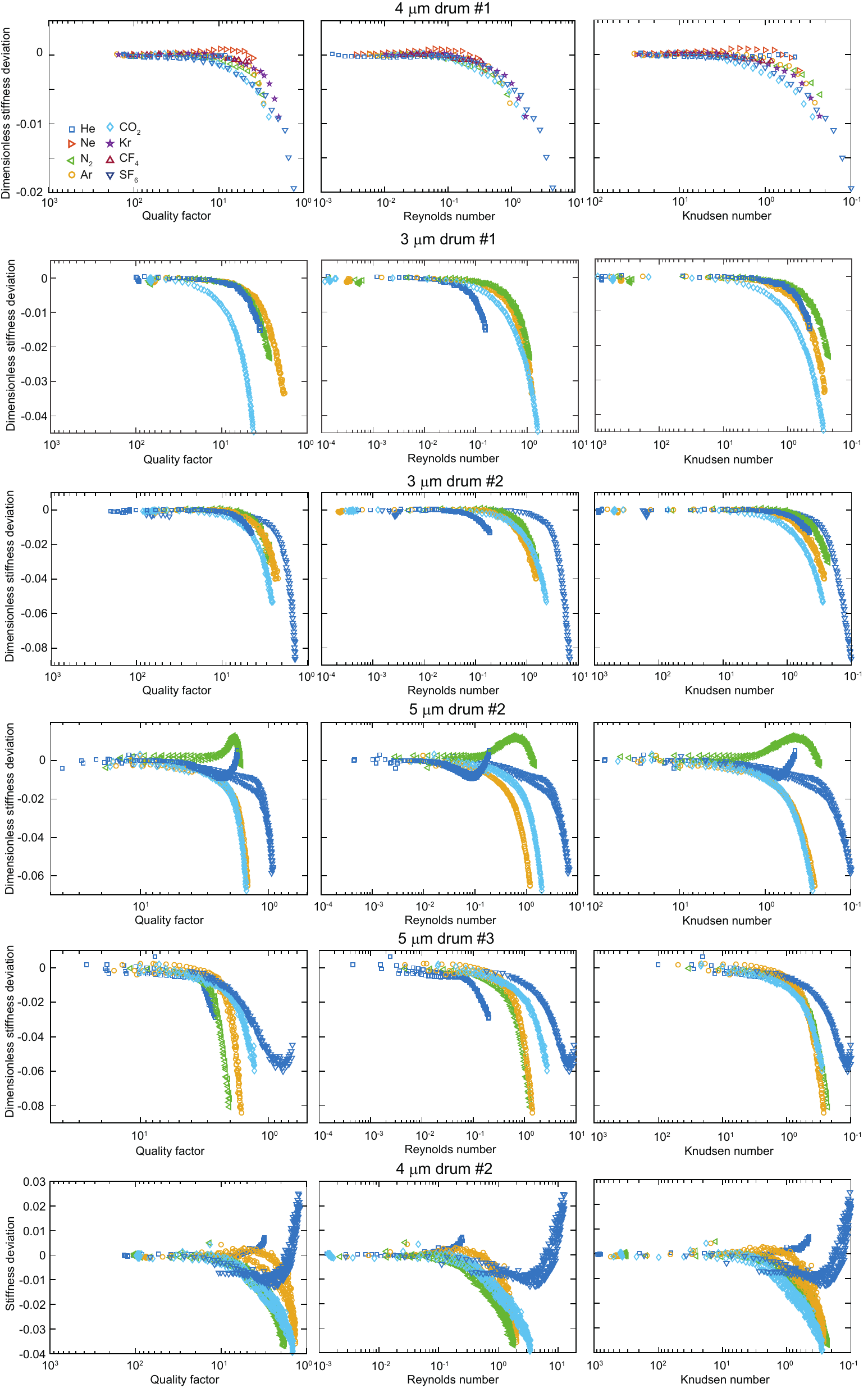}
\caption{Dimensionless number analysis on 6 more samples not shown in the main text.  }
\end{figure*}

\newpage
~
\newpage
~
\newpage
\subsection*{S3: Simulations}
\begin{figure}[h!]
\centering
\includegraphics{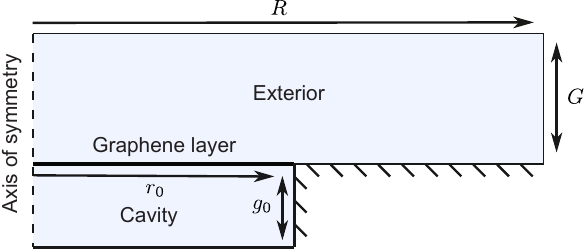}
\caption{ Schematic of domain used for numerical simulations. \label{fig:sqz8}}
\end{figure}

Figure \ref{fig:sqz8} shows a schematic of the domain used in both numerical simulations. For both the continuum and Boltzmann Transport Equation (BTE) simulations the domain was taken to be axisymmetric around the center of the device. For each of the pressures shown in Fig. 6 in the main text, the bounds on the exterior domain, R and G, were increased until a consistent frequency and quality factor was achieved; bounds of at least $R = 3r_0$ and $G = 40g_0$ were used. While for the continuum simulation the drum was taken to be fully enclosed, for the BTE simulation the drum walls were taken to be slightly porous to simulate the presence of the venting channel, with the porosity set to match the proportion of the drum wall occupied by the channel; in the results shown here, this was set at 5\%. As expected, varying this parameter had a negligible effect on the measured frequency.

As discussed in the main text, the continuum simulations applied the eigenfrequency solver of COMSOL \cite{dc2013jpcc,dc2015pof} to the compressible Stokes equation for the gas (with no-slip boundary conditions), and Navier’s equation for the solid. The BTE simulations employed a custom code solving the frequency domain Boltzmann-BGK equation \cite{ladiges2015fdmcA,ladiges2015fdmcB,bhatnagar1954model} for the gas, with diffuse boundary conditions \cite{Bird1994}; this boundary condition naturally includes slip as the gas becomes rarefied. They were coupled with an axisymmetric membrane equation to represent the graphene. For both the continuum and BTE simulations the physical parameters of the graphene were inferred by matching to the frequency measured in a vacuum.

  \begin{figure}[h!]
 \includegraphics{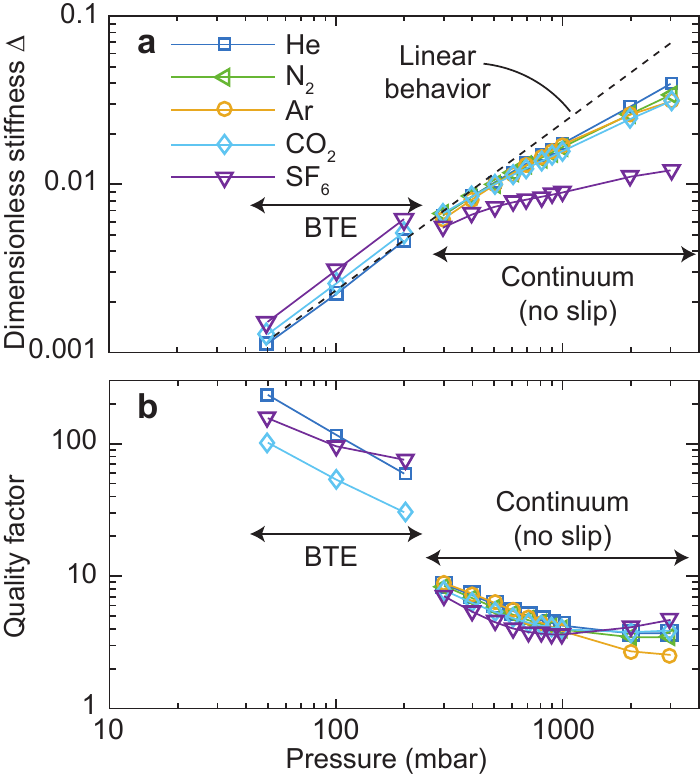}
 \caption{Simulations of resonance frequency and quality factor for a representative squeeze-film pressure sensor. The continuum simulations shown here are used to perform the dimensionless analysis shown in the main text. \textbf{a} Dimensionless frequency shift, showing a clear deviation for sulphur hexafluoride. \textbf{b} Quality factor of resonance, showing the increase at high pressures discussed in the main text.\label{fig:simulations}}
 \end{figure}
Figure \ref{fig:simulations} shows the dimensionless frequency shift and the quality factor from simulations of the previously published 31-layer device. In agreement with the experiments performed here, we find that the continuum simulations predict deviations from the linear stiffness increase as a function of pressure (Fig. \ref{fig:simulations}\textbf{a} ). Sulphur hexafluoride shows a clear deviation from the other gases. The simulated quality factor as a function of pressure in Fig. \ref{fig:simulations} shows a similar weak gas dependence as a function of molecular weight. At high pressures, however, the simulations suggest that the quality factor could increase as a function of pressure. This behaviour has not been observed in experiments, possibly due to the limited pressure range that could be achieved, but is predicted by the single relaxation time model as discussed in the main text. The BTE simulations show a significantly larger quality factor since slip flow is included in this model. The model captures the pressure and gas-dependent trends of the quality factor that are observed in the experimental data but produces an upshift as function of pressure in the case of SF$_6$ that is not observed in the experiments.

\begin{figure}
 \includegraphics{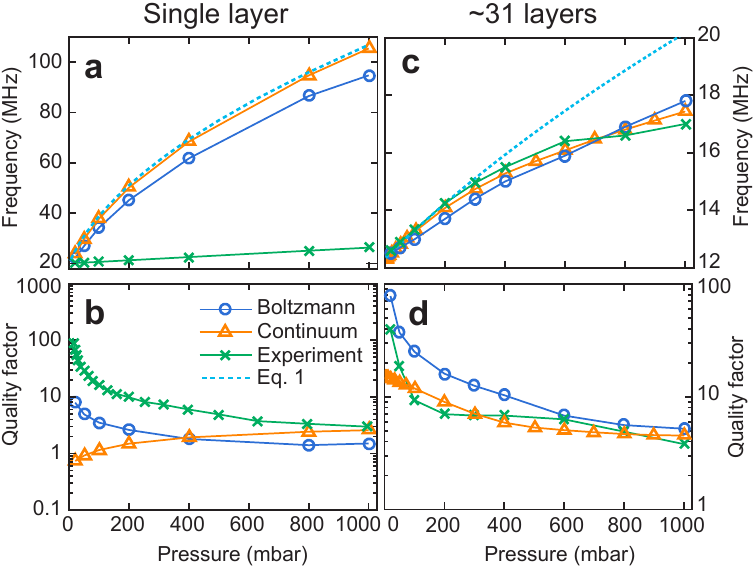}
 \caption{ \textbf{a} Continuum and Boltzmann simulations of the resonance frequency of a 4-micron diameter single-layer graphene drum in a helium environment as a function of pressure in the case of single-layer graphene, compared to Eq. \ref{eq:sqz} and experimental data. The simulations assume the theoretical mass of a clean single-layer graphene membrane. \textbf{b} Quality factor of resonance corresponding to figure \textbf{b}. \textbf{c} Resonance frequency in the case of a 5-micron diameter 31--layer graphene drum in a nitrogen environment and \textbf{d} quality factor of resonance.   \label{fig:modelsingleandmultilayer}}
\end{figure}
Figure \ref{fig:modelsingleandmultilayer}(\textbf{a}-\textbf{b}) shows additional simulation results for a 4-micron diameter single-layer graphene membrane in a helium environment, compared to experimental data. In these simulations we assume the single layer graphene has its theoretical mass of $\num{7.7e-7}$ kg/m$^2$.  In the experiments with single-layer graphene devices a much lower frequency shift is observed compared to the simulations, which may be attributed to additional mass on the single-layer graphene resonator which lowers the resonance frequency and squeeze-number (Fig. \ref{fig:modelsingleandmultilayer}\textbf{a}). Even though the simulations suggest that the quality factor of the resonator will be much lower if the mass is reduced (Fig. \ref{fig:modelsingleandmultilayer}\textbf{b}), the deviations in the resonance frequency are predicted to be small. This is possible because such a clean resonator will operate in the regime where $p_a/g_0 \rho h$ is very large, which corresponds to the high-pressure lines in Fig. 5 in the main text. Figure \ref{fig:modelsingleandmultilayer}(\textbf{c}-\textbf{d}) simulate the previously published 31-layer device \cite{dolleman2015graphene} in a nitrogen environment. This serves as a benchmark for the simulations, but also shows that the deviations from Eq. 1 in the main text can be reproduced in the simulations if the mass of the membrane is increased.

\subsection*{S4: Relation between Q-factor and dimensionless squeeze number}
To describe the graphene squeeze-film pressure sensor, we make a model with a single degree of freedom for the displacement $x$ of the fundamental mode of the graphene membrane:
 \begin{equation}\label{eq:motion2}
 -\omega^2  x + i \omega \frac{ 2}{\tau_0} x + \omega_0^2 x =  \beta \Delta p,
 \end{equation}
 where $\omega$ is the radial frequency, $\tau_0$ the exponential decay time of the membrane in vacuum, $\omega_0$ is the resonance frequency, $\Delta p$ the pressure difference over the membrane due to the squeeze-film effect, $\beta$ is a proportionality constant. 
 For the pressure in the cavity, we take a one-dimensional approximation by assuming the cavity has using a single relaxation time $\tau_g$ (which is the leak time):
 \begin{equation}\label{eq:pressurefield2}
  i \omega \Delta p + \frac{1}{\tau_g} \Delta p = \gamma i \omega x,
 \end{equation}
where $\gamma$ is another proportionality constant. 

To find the free equation of motion we solve Eq. \ref{eq:pressurefield2} for $\Delta p$:
\begin{equation}
\begin{aligned}
  (i \omega + \frac{1}{\tau_g}) \Delta p = \gamma i \omega x,\\
  \Delta p = \frac{i \omega \gamma}{i\omega + \frac{1}{\tau_g}} =\gamma \frac{i \omega \tau_g }{i \omega \tau_g +1},\\
  \Delta p = \gamma\frac{\omega^2 \tau_g^2}{\omega^2 \tau_g^2 +1} + \gamma\frac{ i \omega \tau_g}{\omega^2 \tau_g^2 +1},
\end{aligned}
\end{equation}
This can be substituted into Eq. \ref{eq:motion2}:
\begin{equation}
 -\omega^2  x + i \omega \frac{ 2}{\tau_0} x + \omega_0^2 x =  \beta \gamma \frac{\omega^2 \tau_g^2}{\omega^2 \tau_g^2 +1} + \beta \gamma \frac{ i \omega \tau_g}{\omega^2 \tau_g^2 +1},
\end{equation}
\begin{equation}
 -\omega^2  x + i \omega \left(\frac{ 2}{\tau_0} - \beta \gamma \frac{ \tau_g}{\omega^2 \tau_g^2 +1}\right) x +\left( \omega_0^2 -\beta \gamma \frac{\omega^2 \tau_g^2}{\omega^2 \tau_g^2 +1}\right)x =  0,
\end{equation}
By comparing the case $\omega \tau_g \gg 1$, we note that $\beta \gamma = - \frac{p_a}{g_0 \rho h}$ where $p_a$ is the atmospheric pressure, $g_0$ the gap size and $\rho h$ the membrane's mass per unit square. The stiffness of the system becomes:
\begin{equation}
\omega_f^2 = \omega_0^2 + \frac{p_a}{g_0 \rho h} \frac{\omega^2 \tau_g^2}{\omega^2 \tau_g^2 +1}.
\end{equation}
The damping becomes:
\begin{equation}
\Gamma_f = \frac{ 2}{\tau_0} + \frac{p_a}{g_0 \rho h} \frac{ \tau_g}{\omega^2 \tau_g^2 +1}.
\end{equation}
In the case where $\tau_0 \ll \tau_g$ and close to the resonance frequency:
\begin{equation}
\Gamma_f \approx \frac{p_a}{g_0 \rho h} \frac{ \tau_g}{\omega_f^2 \tau_g^2 +1},
\end{equation}
giving the equation of motion:
\begin{equation}
 -\omega^2  x +  \frac{p_a}{g_0 \rho h}  \frac{\omega_f \tau_g}{\omega_f^2 \tau_g^2 +1} x +\left( \omega_0^2  + \frac{p_a}{g_0 \rho h}  \frac{\omega_f^2 \tau_g^2}{\omega_f^2 \tau_g^2 +1}\right)x =  0.
\end{equation}

The dimensionless squeeze number $\sigma$ compares the timescale of the compression to the timescale of the leakage:
\begin{equation}\label{eq:sigma2}
\sigma = \tau_g \omega_f,
\end{equation}
resulting in:
\begin{equation}
 -\omega^2  x +  i\frac{p_a}{g_0 \rho h}  \frac{\sigma}{\sigma^2 +1} x +\left( \omega_0^2  + \frac{p_a}{g_0 \rho h}  \frac{\sigma^2}{\sigma^2 +1}\right)x =  0,
\end{equation}
The complex eigenvalues of this equation are given by:
\begin{equation}
\omega^2 = \omega_0^2  + \frac{p_a}{g_0 \rho h}  \frac{\sigma^2}{\sigma^2 +1}+   i\frac{p_a}{g_0 \rho h}  \frac{\sigma}{\sigma^2 +1} ,
\end{equation}
\begin{equation}
\begin{aligned}
\omega = \sqrt[\leftroot{0}\uproot{2}4]{ \left(\frac{p_a}{g_0 \rho h}  \frac{\sigma(\sigma+1)}{\sigma^2 +1}  \right)^2 + \left(\frac{p_a}{g_0 \rho h}  \frac{\sigma}{\sigma^2 +1}\right)^2 } \times \\
\cos \left(\frac{1}{2} \arg\left(\omega_0^2  + \frac{p_a}{g_0 \rho h}  \frac{\sigma^2}{\sigma^2 +1}+   i\frac{p_a}{g_0 \rho h}  \frac{\sigma}{\sigma^2 +1} \right) \right) +\\
i \sqrt[\leftroot{0}\uproot{2}4]{ \left(\frac{p_a}{g_0 \rho h}  \frac{\sigma(\sigma+1)}{\sigma^2 +1}  \right)^2 + \left(\frac{p_a}{g_0 \rho h}  \frac{\sigma}{\sigma^2 +1}\right)^2 } \times \\
\sin \left(\frac{1}{2} \arg\left(\omega_0^2  + \frac{p_a}{g_0 \rho h}  \frac{\sigma^2}{\sigma^2 +1}+   i\frac{p_a}{g_0 \rho h}  \frac{\sigma}{\sigma^2 +1} \right) \right).
\end{aligned}
\end{equation}
Using $Q_f = \mathcal{R}(\omega)/2\mathcal{I}(\omega)$:
\begin{equation}
\begin{aligned}\label{eq:Qf2}
Q_f^2 = \frac{\cos^2 \left(\frac{1}{2} \arg\left(\omega_0^2  + \frac{p_a}{g_0 \rho h}  \frac{\sigma^2}{\sigma^2 +1}+   i\frac{p_a}{g_0 \rho h}  \frac{\sigma}{\sigma^2 +1} \right) \right) }{ 4\sin^2 \left(\frac{1}{2} \arg\left(\omega_0^2  + \frac{p_a}{g_0 \rho h}  \frac{\sigma^2}{\sigma^2 +1}+   i\frac{p_a}{g_0 \rho h}  \frac{\sigma}{\sigma^2 +1} \right) \right) } \\
Q_f = \frac{1}{2} \cot \left(\frac{1}{2} \arctan\left( \frac{\frac{p_a}{g_0 \rho h}  \frac{\sigma}{\sigma^2 +1}}{\omega_0^2  + \frac{p_a}{g_0 \rho h}  \frac{\sigma^2}{\sigma^2 +1}} \right)\right) \\
Q_f = \frac{1}{2} \cot \left(\frac{1}{2} \arctan\left( \frac{ \frac{\sigma}{\sigma^2 +1}}{\frac{\omega_0^2}{\omega_{\mathrm{sqz}}^2} +   \frac{\sigma^2}{\sigma^2 +1}} \right)\right),
\end{aligned}
\end{equation}
\begin{equation}\label{eq:QfSI1}
Q_f = \frac{\sqrt{\xi^2 + 1} + 1}{2 \xi},
\end{equation}
where:
\begin{equation}\label{eq:QfSI2}
\xi = \frac{\sigma }{\frac{\omega_0^2}{\omega_{\rm sqz}^2} (\sigma^2 + 1) + \sigma^2 }.
\end{equation}

\section*{S5: Relation between relative stiffness deviation and Q factor}
In the main text, we define the relative stiffness deviation as:
\begin{equation}\label{eq:deltarel2}
\Delta_{\rm rel} = \frac{\Delta - \Delta_{\rm lin}}{\Delta_{\rm lin}}
\end{equation}
To calculate this, we need the resonance frequency, which is given by the absolute value of $\omega$ in Eq. 7 in the main text:
\begin{equation}
|\omega| = \sqrt[4]{\frac{\omega_0^4 + \sigma^2(\omega_0^2 + \frac{p_a}{g_0 \rho h})^2}{\sigma^2 +1}}
\end{equation}
If we assume $\Delta_{\rm lin}$ is perfectly described by Eq. 1 from the main text, and that $\Delta$ is given by the real part of Eq. 7, we obtain:
\begin{equation}
\Delta_{\mathrm{rel}} = \frac{g_0 \rho h}{p_a} \sqrt{\frac{\omega_0^4 + \sigma^2(\omega_0^2 + \frac{p_a}{g_0 \rho h})^2}{\sigma^2 +1}} - \frac{\omega_0^2 g_0 \rho h}{p_a} - 1
\end{equation}
Note, that by using $\Delta_{\mathrm{rel}}$, instead of $\Delta - \Delta_{\mathrm{lin}}$, the theoretical value of $\rho h$ and $p_{\rm ref}$ used for the normalization in Eq. 2 in the main text have dropped out. $\Delta_{\rm rel}$ now only depends on the real pressure and mass of the device. 
Next, we invert this equation to obtain an expression for $\sigma$, keeping only the positive solution:
\begin{equation}
\sigma = \frac{i \sqrt{\Delta_{\rm rel}+1} \sqrt{2 \omega_0^2 + \frac{p_a}{g_0 \rho h} \Delta_{\rm rel} + \frac{p_a}{g_0 \rho h}}}{\sqrt{\Delta_{\rm rel}} \sqrt{2 \omega_0^2 + \frac{p_a}{g_0 \rho h} \Delta_{\rm rel} + 2 \frac{p_a}{g_0 \rho h}}}
\end{equation}
This can then be evaluated at different values of $\Delta_{\rm rel}$ and substituted into Eqs. \ref{eq:QfSI1}--\ref{eq:QfSI2} to produce the red lines in Fig. 5 in the main text. 

\end{document}